# Structural, electronic and bonding properties of zeolite Sn-Beta: A periodic density functional theory study

**Sharan Shetty,[a] Sourav Pal,[a,*] Dilip G. Kanhere,[b] Annick Goursot[c]**

[a]Theoretical Chemistry Group, Physical Chemistry Division, National Chemical Laboratory, Pune-411008

[b]Centre for Modeling and Simulation and Department of Physics, University of Pune, Pune-411008, India

[c]Ecole de Chemie e Montpellier, UMR 5618 CNRS Ecole de Chemie, 8, rue de l'Ecole, Normale 34296, Montpellier, Cedex 5 France.

**Abstract:**

The structural, electronic and the bonding properties of the Sn-BEA are investigated by using the periodic density functional theory. Each of the 9 different T-sites in the BEA were substituted by the Sn atom and all the 9 geometries were completely optimized using the plane wave basis set in conjunction with the ultra-soft pseudopotential. On the basis of the structural and the electronic properties, it has been demonstrated that the substitution of the Sn atom in the BEA framework is an endothermic process and hence the incorporation of the Sn in the BEA is limited. The lowest unoccupied molecular orbitals (LUMO) energies have been used to characterize the Lewis acidity of each T-site. On the basis of the relative cohesive energy and the LUMO energy, T2 site is shown to be the most favorable site for the substitution of Sn atom in the BEA framework.

## I. Introduction

Zeolite beta (BEA) was synthesized in 1967 and showed high catalytic activity.[1] The structure of BEA was only recently determined, because the crystals of BEA always contains severe structure faulting and hence shows strong diffuse scattering in diffraction patterns. In 1988 Newsam *et. al*,[2] succeeded in solving the structure mainly with the use of electron microscopy. They showed that the structure of BEA consists of an inter-grown hybrid of two distinct polytypic series of layers viz. polymorph A and B. Both the polymorphs have 3D network of 12-ring pores. The polymorph grows as two dimensional sheets and the sheets randomly alternate between the two. Very recently polymorph C has been proposed by Corma *et*

*al*.[3] BEA has two mutually perpendicular straight channels each with a cross section of 0.76 x 0.64 nm along the a and b direction and a helical channel of 0.55 x 0.55 nm along the c-axis. BEA is of great industrial interest because of its high acidity and larger pore size.[4] BEA has been successfully used for acid catalyzed reactions,[5] catalytic cracking,[6] aromatic and aliphatic alkylation.[7] It has been shown that the acidity of BEA can be tuned by the incorporation of trivalent and tetravalent atoms (B, Al, V, Ti, Sn, Cr, Fe) into the framework positions of BEA.[8-13] The isomorphous substitution creates various Bronsted and Lewis acid sites in BEA. It has been shown in the earlier studies that the Bronsted acid sites are present, both in the internal as well as on the external surface.[13] However, this is not true for the Lewis acid sites. There are predominantly present in the framework.[13] The nature of the acid sites created by the isomorphous substitution plays a crucial role in understanding some important oxidation and reduction reactions.[11-15] Interestingly NMR studies have been used to characterize the nature and the acidity of various active sites in BEA.[16,17]

In recent years it has been shown that the isomorphous substitution of Si by Ti into the zeolite framework provides a useful catalyst for the oxidation of organic compounds. This is attributed to the strong Lewis acidity of Ti.[11-15, 18] It is known that the Ti T-sites, which act as Lewis acid, are also the active sites and are located inside the BEA framework in a tetrahedral coordinated position. Many experimental methods such as IR spectroscopy, XRD, thermal analysis, XPS, EXAFS-XANES have been employed to study the characterization of Ti sites in BEA. One of the major advantage of the Ti-BEA over titanium silicate (TS-1) is its pore size which allows an easy access to larger reactant molecules.[12] Apart from the different pore size the catalytic activities of Ti-BEA and TS-1 are quite different. Jansen *et al* have shown that the Ti-BEA acts as an active catalyst for Meerwin-Ponndorf-Verley (MPV) reduction of aldehydes and ketones and Oppenauer's (O) oxidation of alchols (MPVO reactions).[13] Adam *et al* have carried out the catalaytic oxidation of silanes to silanols in the presence of $H_2O_2$ by Ti-BEA.[20]

Recently, Sn-BEA was shown to have better catalytic activity than Ti-BEA.[8] Al-free-Sn BEA was first synthesized by Mal and Ramaswamy whopredicted that the Sn to be tetrahedrally coordinated.[21] Corma *et al*, for the first time showed that the Sn-BEA acts as an efficient catalyst for the Baeyer-Villiger oxidation reaction in the presence of $H_2O_2$.[8] They further showed the probable reaction mechanism of the Baeyer-Villiger oxidation reaction, in which the carbonyl group of the ketone is initially activated followed by a reaction with the non-activated $H_2O_2$, unlike the Ti-BEA zeolite.[8] Later, it was also shown by Corma *et al,* that Sn-BEA acts as a better catalyst than the Ti-BEA for the MPVO reaction.[22, 23] The increase in the activity of BEA by the substitution of Sn can be rationalized by the higher atomic size and electro-negativity than the Ti atom giving rise to stronger Lewis acid sites. This shows that the combined property of large pore dimension and high Lewis acidity of Sn-BEA makes it a highly active stereo-selective catalyst for many oxidation and reduction reactions. The $Sn^{119}$ MAS NMR spectrum of Sn-BEA confirmed that the Sn has tetrahedral coordination in contrast to the octahedral coordination.[22-24] This was also attributed to the Sn active sites present within the framework, but not on the external surface in the form of $SnO_2$.[24] It has been shown that the substitution of Si by a Ti atom in the zeolite framework always results in an increase of the cell volume.[25] It is known that the crystallographically inequivalent T-sites will have different activity and shape selectivity due to the differences in the topological environment around the T-sites. Hence, in the isomorphic substituted zeolites such as Sn-BEA and Ti-BEA, it is important to understand the nature of the active sites and the structural quantification of these active sites. Experimental techniques such as X-ray, MAS NMR and IR have been adopted to understand the coordination of the active sites in the zeolite and the interaction of these sites with the organic molecules.[23, 24, 26,27] However, in zeolites such as Sn-BEA or Ti-BEA, where the concentration of the Sn or Ti is low in the framework, it becomes difficult to obtain the structural features of the local active sites using these experiments.[24,26]

Various quantum mechanical methods have been implemented to study the structural and electronic properties of the active sites in zeolites. Finite or cluster models of an active site, cut out of the zeolite crystal have been used for theoretical investigation, where the dangling bonds of the cluster are saturated by hydrogen atoms. The reviews by Sauer and coworkers may be referred for detailed study on the cluster models of zeolites.[28,29] Advantage of using the cluster model is that, it presents a simple model of zeolites and is computationally cheap. It is also a better model for representing the active sites on the surface. However, cluster models neglect the effect of long range interactions and some artificial states are introduced due to the atoms lying at the boundary of the truncated fragment. Periodic methods are the only way to overcome all these problems, which include the long range electrostatic interactions and the effect of the zeolite cage. Studies on the comparison of clusters versus the periodic calculations have been carried out in past.[29,30]

Sastre and Corma have carried out cluster calculations on the Ti-BEA and TS-1 using ab-initio Hartree-Fock and density functional theory.[31] On the basis of LUMO energies they characterized the acidity of these two zeolites and proved that Ti-BEA is more Lewis acidic than TS-1.[31] Dimitrova and Popova have done a cluster study of Al, B, Ti, and V incorporated BEA and further studied their interaction with the peroxogroup (O-O-H).[32] They showed that the incorporation of Ti is favorable than the other atoms and Ti increases the oxidizing power of the peroxo group. Zicovich-Wilson and Dovesi carried out periodic Hartree-Fock calculations on Ti-containing zeolites such as SOD, CHA and alpha-quartz (QUA).[25] Interestingly, they showed that the substitution of Si atom by Ti atom in a zeolite is an endothermic process when evaluated with respect to pure silicozeolite. They also proved that the incorporation of Ti within the zeolite framework is thermodynamically less favored than the formation of extra-framework $TiO_2$ clusters. This explains the difficulty of synthesizing high Ti content zeolites.[25] Very recently, Damin *et al,* have studied the interaction of Ti-CHA with various molecules such as $NH_3$, $H_2O$, $H_2CO$ and $CH_3CN$ using a periodic approach.[33] Moreover, there has been a lot of studies on other zeolites using periodic DFT.[34] Recently, Rozanska *et al* have used the periodic DFT approach to study the chemisorption of several organic molecules in zeolites.[35]

As discussed above, the Sn centers are catalytically very important and act as stronger Lewis acidic sites during the oxidation and reduction reactions. Thus, it is necessary to obtain the information on the structural and electronic properties of Sn-BEA. Motivated by this, in the present work, we wish to examine the effect of the incorporation of Sn in BEA zeolite using periodic DFT. The paper is organized as follows: In section II we present the computational methods used. In section III we present the results and discussion. Section IV presents the important conclusions of the paper.

## II. Computational Method

All the calculations presented in this paper has been performed by periodic DFT implemented in Vienna *ab initio* simulation package (VASP) code.[36] The instantaneous electronic ground state is calculated by solving the Kohn-Sham equation which calculates the current wavefunctions using a residual vector and are supposed to converge if the value of the residual vector is less than some specified cut-off value. The present method uses the plane wave basis set in conjunction with the ultra-soft Vanderbilt pseudopotentials.[37] Advantage of using this computationally efficient scheme is that it permits the use of Fast Fourier transform technique. Plane waves are delocalized basis set and hence are free from the basis set superposition errors and allow accurate calculation of total energy of different atomic arrangements. The exchange correlation functional is expressed within the generalized gradient approximation (GGA) with the Perdew-Wang 91 functional.[38] The Brillouin zone sampling was restricted to the gamma point.

Structural relaxation of the coordinates of the BEA and the Sn-BEA have been performed in two steps. Initially, the conjugate gradient method has been employed to optimize the structures during which the cell shape of the unit cell has been fully relaxed by keeping the volume fixed. This was done until the forces on the atoms were less than 0.1 eV/A. In the next step the optimized structure obtained from the conjugate gradient was used as the starting geometry and was re-optimized using the quasi-Newton method unless the forces on the atoms were less than 0.06 eV/A. BEA has 9 inequivalent x-ray crystallographically defined T-sites.[31] During the Sn-BEA unit cell optimization, all these 9 T-sites were substituted one by one by Sn atom such that only one Sn atom is present per unit cell i.e. Si/Sn=63/1. The unit cell was also optimized with two Sn atoms per unit cell as described above i.e. Si/Sn=62/2.

### III. Results and Discussion
**(A.) Structure**

The optimized structural parameters of Sn-BEA for all the 9 T-sites (T1 to T9) are given in Table. 1. As discussed in the introduction, BEA has 9 inequivalent X-ray crystallographically defined T-sites (Fig. 1). There are basically 192 atoms in the unit cell with 64 Si atoms and 128 O atoms. The distribution of these 64 Si atoms are as follows; There are 8 Si atoms placed at the T1 to T6 and T8 positions, while 4 Si atoms are placed at the T7 and T8 positions. Only the average Sn-O, Sn-Si bond lengths and Sn-O-Si bond angles are presented. However, the optimized average, Si-O bond distance and the Si-O-Si bond angles of BEA are 1.612±0.002 Å and 149.25±1.5 deg respectively, which are in good agreement with the earlier studies.[11] As expected, after the substitution of the Sn atom in the BEA the average Sn-O bond distance increases to 1.912±0.002 Å and the bond angles range from 137 to 147 deg. This shows that, after the substitution of Sn in the BEA framework the Sn-O bond distance increases by about 0.3 Å and the bond angles decreases by about 2 to 10 deg. Although there is a decrease in the bond angle of the Sn-O-Si, the Sn-Si distance is more than the Si-Si distance in BEA, which is due to the increase in the Sn-O distance. It has been already confirmed by the experimental studies such as the $^{119}$Sn MAS NMR, that the Sn in the Sn-BEA is situated in the framework and in a tetrahedral coordination.[8,22-24] The O-Sn-O angles have also been calculated and it was seen that the angles significantly deviate from the tetrahedral value. These kind of O-T-O angle deviations from the tetrahedral values are also reported for the titanosilica zeolite models.[25] However, the average O-Sn-O bond angles tend to remain close to ~ 109.5 deg. Unfortunately, there are no earlier theoretical studies on Sn-BEA to compare with the results presented in this work. The change in the bonding due to the distortion in the the local Sn-site of Sn-BEA is discussed in the next section. We observe from Table. I, that the T2 and the T8 positions have the shortest Sn-O bond lengths of 1.909 and 1.908 Å respectively. However, the bond angle of the Sn-O-Si at the T2 position is the largest with 144.2 deg. The Sn-Si distance is the largest for T2 position, which can be attributed to the larger Sn-O-Si bond angle. The T4 position has the largest Sn-O bond distance of about 1.917 Å and a shortest Sn-O-Si bond angle of 136.0 deg. The change in the structural parameters of each T site can drastically affect the energetics of the zeolite. Hence, the stability and the reactivity of each site would be different.

It should be noted that the present structural results should not be compared with the cluster models of Sn-BEA, due to the fact that the cluster models do not account the long range interaction. Moreover, the Sn sites are located within the zeolite framework and cluster models do not consider the effect of the cage atoms.[29]

**(B.) Relative cohesive energy and stability.**

In this subsection we discuss the relative cohesive energies and hence the stabilities of all the 9 substituted Sn-sites. The cohesive energies for all the 9 T sites are given in Table. II. The cohesive energy is the difference between the energy of the bulk (solid) at equilibrium and the energy of the constituent atoms in there ground state combined. The cohesive energy is defined as

$$E_{coh} = E_{solid} - \sum_{i} E_i$$

where, i represents the individual atoms that constitute the solid.
Higher the cohesive energy of the solid more stable it is. The cohesive energy of BEA is -

1527.9026 eV. From Table. II, we see that the cohesive energy of Sn-BEA ranges between -1521.32 to -1521.68 which is about ~6 eV lower than the BEA. This explains the fact that the substitution of Sn in the BEA framework decreases the cohesive energy. Hence, substitution of the Sn atom in the BEA framework is an endothermic process i.e. thermodynamically, it is less favored. This clearly indicates that more Sn cannot be incorporated during the synthesis of Sn-BEA. To confirm this, we performed an optimization of Sn-BEA with 2 Sn atoms in the unit cell and it was seen that the cohesive energy is decreased by about 6 eV (138 kcal/mol). This may be the reason for the decrease in the turnover number and selectivity of cyclohexanone in the Baeyer-Villiger reaction, as proved by the experiments.[8,22-24]

Among the 9 T-sites, the T2 site has the highest cohesive energy (Table. II) and hence shows the most stable site for the substitution of Sn atom in Sn-BEA. This can also be attributed to the shorter Sn-O distance and longer Sn-O-Si bond angle. The next most stable site is the T8 site which has ~1.5 kcal/mol less cohesive energy than the T2 site. The most unstable site is the T9 site which has ~8.23 kcal/mol less energy than the T2 site (Table. II).

**(C.) LUMO energies**

It has been known that, lower the LUMO energy of a system, higher is its ability to gain electron density and hence has a higher Lewis acidity. Sastre and Corma used the LUMO energies to characterize the Lewis acidity of the Ti sites in Ti-BEA and TS-1.[31] They showed that the average LUMO energy of the Ti sites in Ti-beta is lower than the TS-1 and hence Ti-beta was shown to have a higher Lewis acidity than TS-1. On this basis, in the present work we use the LUMO energies to discuss the Lewis acidity of all the 9 T sites in Sn-BEA. The LUMO energies of all the 9 T-sites are given in Table. II. The results show that the T1 and the T2 sites in Sn-BEA have the lowest LUMO energies and should be the most probable Lewis acid sites. However, T1 site has a lower cohesive energy than the T2 site as discussed in the earlier section. Hence, on the account of high cohesive energy and low LUMO energy, the T2 site would be the most probable site for the substitution for the Sn atom in Sn-BEA.

The LUMO energy calculated for BEA is 2.643 eV, which is about 1.2 to 1.3 eV higher than the Sn-BEA (Table. III). This clearly shows that the substitution of Sn in BEA drastically increases the Lewis acidity.

**(D.) Bonding**

In this subsection we focus on the nature of bonding in the Sn-BEA and compare this to the same in BEA. In Fig. 2 and 3 we plot the highest occupied molecular orbital (HOMO) and the LUMO isodensities respectively of the BEA at one third of the maximum isosurface value. We can clearly see that the HOMO of BEA indicates a pi-orbital on the oxygen atom. However, the LUMO isodensities of BEA (Fig. 3) clearly show lone pair of electrons capped on the oxygen atoms.

Fig. 4 and 5 show the respective HOMO and LUMO isodensities of the Sn-BEA at one third of the maximum isosurface value. The HOMO isodensity of Sn-BEA is different from the Si-BEA which does not show any pi-orbital on oxygen atoms attached to the Sn atom (Fig. 4), as it was seen in Si-BEA. Interestingly, the LUMO isodensity of Sn-BEA shows that the lone pair of electrons on the oxygen atoms coordinated to the Sn atom are polarized by the Sn atom (Fig. 5), which is not seen in the LUMO of Si-BEA. This explains the fact that the Sn atom has more ability to polarize the electron density of the oxygen atoms. This can be attributed to the high Lewis acidity of Sn atom and explains polar bond of Sn with oxygen atoms.

**IV. Conclusion**

In the present work we have discussed the structure, bonding and the acidity of the Sn substituted BEA using periodic density functional approach. The results demonstrate that the incorporation of Sn in the BEA framework decreases the cohesive energy and is an endothermic process. Hence, it is clear that the incorporation of Sn in the BEA is limited. This would be the reason for the decrease in the turn-over number as the Sn content is increased in BEA during the Baeyer-Villiger oxidation reaction. The structural parameters, as expected, show an increase in the Sn-O bond lengths compared to the Si-O bond lengths in BEA. Among the 9 T-sites, T2 site proves to be the most stable site for the substitution of Sn in the BEA framework, which is due to the higher cohesive energy compared to the other T sites. Moreover, the T2 site shows a higher Lewis acidic site compared to the other T-sites. The bonding analysis show that the Sn atom polarizes the lone pair of electrons of the oxygen atoms, which can be attributed to its higher electronegativity.

The present theoretical study gives an insight into the structural and the electronic properties of the T sites in Sn-BEA which is otherwise difficult by using the experimental techniques. Further work on the effect of the solvent molecules on the Sn-sites is still in progress.

**Acknowledgement:**

S. Shetty, S. Pal and A. Goursot gratefully acknowledge the Indo-French Center for the Promotion of Advance Research (IFCPAR) (Project. No. 2605-2), New Delhi, India, for financial assistance.

[*]To whom correspondence should be addressed: Email: pal@ems.ncl.res.in

**References:**

1. Wadlinger, R. L.; Kerr, G. T.; Rosinski, E. J. U.S. Pat. 3 308 069, 1967

2. Newsam, J. M.; Treacy, M. M. J.; Koestsier, W. T.; de Gruyter, C. B. Proc. R. Soc. London 1988, A420, 375.

3. (a) Corma, A; Navarro, M. T.; Rey, F.; Rius, J.; Valencia, S. *Angew. Chem. Int. Ed.* **2001**, *40*, 2277. (b) Corma, A.; Navarro, M. T.; Rey, F.; Valencia, S. *Chem. Commun.* **2001**, 1486. (c) Ohusna, T.; Liu, Z.; Terasaki, O.; Hiraga, K.; Camblor, M. A. J. Phys. Chem. B 2002, 106, 5673.

4. Martens, J. A.; Perez-Pariente, J.; Sastre, E.; Corma, A,; Jacobs, P. A. Appl. Catal. 1988,45,85. (3) Ratnasamy, P.; Bhat, R. N.; Pokhriyal, S.K.; Hagde, S. G.; Kumar, R. J. Catal. 1989,119,65. (20) Reddy, K. S. N.; Eapen, M. J.; Soni, H. S.; Shiralkar, P. V. J. Phys. Chem. 1992,96,7923.

5. Bellusi, G.; Pazzuconi, G.; Perego, C.; Girotti, G.; Terzoni, G. J. Catal. 1995, 157, 227.

6. Boretto, L.; Camblor, M. A.; Corma, A.; Perez-Pariente, J. Appl. Catal. 1992,82,37.

7. A.J. Hoefnagel and H. van Bekkum, Appl. Catal. A, 97 (1993) 87. (b) K.P. de Jong, C.M.A.M. Mesters, D.G.R. Peferoen, P.T.M. van Brugge and C. de Groot, Chem. Eng. Sci., 51 (1996) 2053

8. Corma, A.; Nemeth, L. T.; Renz, M.;Valencia, S. Nature. 2002, 412, 423.

9. Camblor, M. A.; Corma, A.; Martinez, A.; Perez-Pariente, J. J. Chem. Soc. Chem. Commun. 1992, 589.

10. Sen, T.; Chatterjee, M.; Sivsanker, S. J. Chem Soc. Chem. Commun. 1995, 207.

11. van der Waal, J. C.; van Bekkum, H. J. Mol. Catal. A. 1997, 124, 137

12. Camblor, M. A.; Corma, A.; Perez-Pariente. J. Zeolites. 1993, 13, 82

13. Jansen, J. C.; Creyghton. E. J.; Njo, S. L.; van Koningsveld, H.; van Bekkum, H. Catal. Today. 1997, 38, 205.

14. Corma, A.; Camblor, M. H.; Esteve, P.; Martinez, A.; Perez-Pariente, J. J. Catal. 1994, 145, 151.

15. Corma, A.; Esteve, P.; Martinez, A.; Valencia, S. J. Catal. 1995, 152, 18

16. de Menorval, L. C.; Buckermann, W.; Figueras, F.; Fajula, F. J. Phys. Chem. 1996, 100, 465

17. Valerio, G.; Goursot, A.; Vetrivel, R.; Malkina, O.; Malkina, V.; Salahub, D. R. J. Am. Chem. Soc. 1998, 120, 11426.

18. Corma, A.; Esteve, P.; Martinez, A. Zeolites. 1996, 161, 11.

19. Carati, A.; Flego, C.; Previde Massara, E.; Millini, R.; Carluccio, L.; Parker Jr, N. D.;

Bellussi, G. Microporous and Mesoporous Mater. 1999, 30, 137.

20. Adam, W.; Garcia, H.; Mitchell, C. M.; Saha-Mollera, C. R.; Weichold, O. Chem. Commun. 1998, 2609

21. Mal. N. K.; Ramaswamy, A. V. Chem. Commun, 1997, 425.

22. Corma, A.; Domine, M. E.; Nemeth, L.; Valencia, S.; J. Am. Chem. Soc. 2002, 124, 3194

23. Corma, A.; Domine, M. E.; Valencia, S. J. Catal. 2003, 215, 3194

24. Renz, M.; Blasco, T.; Corma, A.; Formes, V.; Jensen, R.; Nemeth, L. Chem. Eur. J. 2002, 8, 4708

25. Zicovich-Wilson, C. M.; Dovesi, R. J. Phys. Chem. B. 1998, 102, 1411

26. Blasco T.; Camblor, M. A.; Corma, A.; Perez-Pariente, J. J. Am. Chem. Soc. 1993, 115, 11806

27. Blasco, T.; Camblor, M. A.; Corma, A.; Esteve, P.; Guil, J. M.; Martinez, A.; Perdigon-Melon, J. A.; Valencia, S. J. Phys. Chem. B. 1998, 102, 75.

28. Sauer, J. Chem. Rev. 1989, 89, 199

29. Sauer, J.; Ugliengo, P.; Garrone, E.; Saunders, V. R.; Chem. Rev. 1994, 94, 2095

30. Hill, J-R.; Freeman, C. M.; Delley, B. J. Phys. Chem. A 1999, 103, 3772-3777

31. Sastre, G.; Corma, A. Chem. Phys. Lett. 1999, 302, 447.

32. Dimitrova, R.; Popova, M. Mol. Eng. 1999, 8, 471

33. Damin, A.; Bordiga, S.; Zecchina, A.; Doll, K.; Lamberti, C. J. Chem. Phys. 2003, 118, 10183.

34. (a) Shah. R.; Gale, J. D.; Payne, M. C. J. Phys. Chem. 1996, 100, 11688. (b) Demuth, T.; Hafner, J.; Benco, L.; Toulhoat, H. J. Phys. Chem. B 2000, 104, 4593.

35. Rozanska, X.; Demuth, Th.; Hutschka, F.; Hafner, J.; van Santen, R. A. J. Phys. Chem. B 2002, 106, 3248-3254

36. (a) Kresse, G.; Hafner, J. Phys. Rev. B 1994, 49, 14251. (b) Kresse, G.; Furthmuller, J. Computat. Mater. Sci. 1996, 6, 15

37. Vanderbilt, D. Phys. ReV. B 1990, 41, 7892.

38. Perdew, J. P.; Wang, Y. Phys. ReV. B 1992, 45, 13244.

Table. I: Average Bond lengths (A) and bond angles of Sn-O and Sn-O-Si and the next nearest neighbor Sn-Si distance (A) of the optimized structures for the 9 different T-sites in Sn-BEA.

| Sn-Sites | Sn-O | Sn-O-Si | Sn-Si |
| --- | --- | --- | --- |
| T1 | 1.911 | 143.5 | 3.336 |
| T2 | 1.909 | 144.2 | 3.341 |
| T3 | 1.910 | 140.6 | 3.241 |
| T4 | 1.917 | 136.0 | 3.281 |
| T5 | 1.913 | 142.2 | 3.297 |
| T6 | 1.910 | 141.2 | 3.297 |
| T7 | 1.911 | 140.6 | 3.282 |
| T8 | 1.908 | 140.0 | 3.282 |
| T9 | 1.912 | 137.8 | 3.270 |

Table. II Cohesive energies and the lowest unoccupied molecular orbital (LUMO) energies of the 9 different T-sites

| Sn-Sites | Cohesive Energy (eV) | LUMO Energies (eV) |
|---|---|---|
| T1 | -1521.3871 | 1.333 |
| T2 | -1521.6818 | 1.366 |
| T3 | -1521.4687 | 1.557 |
| T4 | -1521.5232 | 1.421 |
| T5 | -1521.4052 | 1.450 |
| T6 | -1521.4316 | 1.426 |
| T7 | -1521.4571 | 1.419 |
| T8 | -1521.6215 | 1.497 |
| T9 | -1521.3239 | 1.506 |

**Figure Captions.**

Figure.1: Unit cell of BEA consisting of 192 atoms. The 9 active sites are shown by the Sn atoms (yellow spheres). The brown and the red spheres are the Si and the O atoms respectively.

Figure. 2 : Isosurface of HOMO of BEA. Blue and red spheres are the Si and O atoms respectively

Figure. 3 : Isosurface of LUMO of BEA. Blue and red spheres are the Si and O atoms respectively

Figure. 4: Isosurface of HOMO of Sn-BEA. Green and red spheres are the Si and O atoms respectively and the Sn atom is shown by blue sphere.

Figure. 5 : Isosurface of LUMO of Sn-BEA. Green and red spheres are the Si and O atoms respectively and the Sn atom is shown by blue sphere.

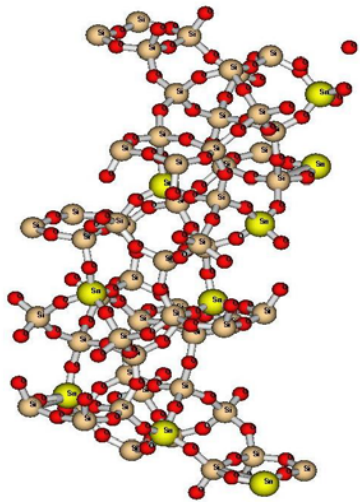

Figure. 1

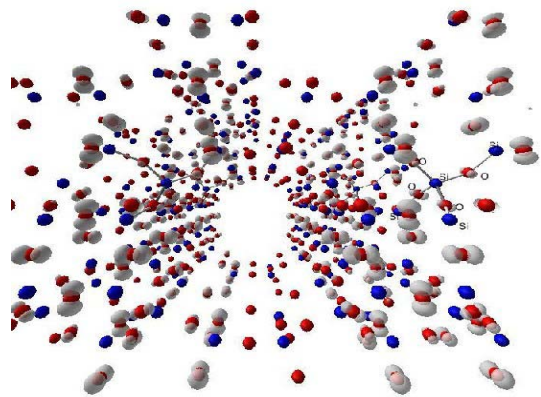

Figure. 2

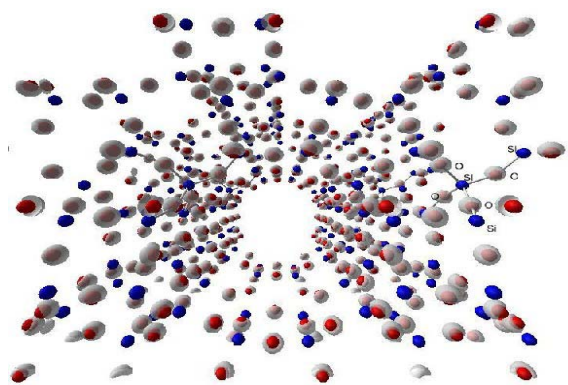

Figure. 3

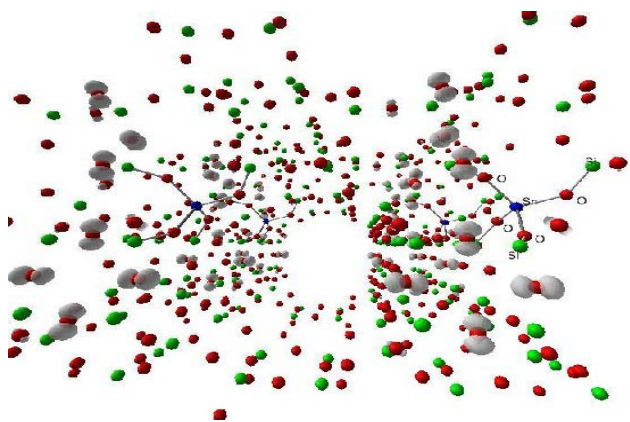

Figure. 4

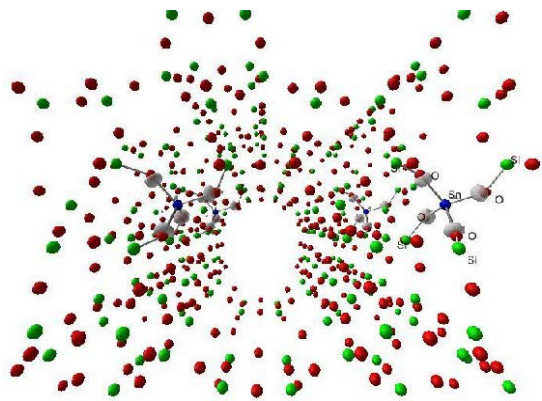

Figure. 5